\documentclass[conference]{IEEEtran}
\IEEEoverridecommandlockouts
\usepackage{cite}
\usepackage{amsmath,amssymb,amsfonts}
\usepackage{algorithm}
\usepackage{graphicx}
\usepackage{textcomp}
\usepackage{xcolor}
\usepackage{algorithm}
\usepackage{algpseudocode}
\usepackage[colorlinks,
            linkcolor=red,
            anchorcolor=blue,
            citecolor=green
            ]{hyperref}
\usepackage{multirow}

\newcommand{\tabincell}[2]{\begin{tabular}{@{}#1@{}}#2\end{tabular}}

\def\BibTeX{{\rm B\kern-.05em{\sc i\kern-.025em b}\kern-.08em
    T\kern-.1667em\lower.7ex\hbox{E}\kern-.125emX}}
\begin{document}

\title{TSEML: A task-specific embedding-based method for few-shot classification of cancer molecular subtypes
}

\author{
    \IEEEauthorblockN{Ran Su$^{a}$, Rui Shi$^b$, Hui Cui$^c$, Ping Xuan$^d$, Chengyan Fang$^a$, Xikang Feng$^b$, and Qiangguo Jin$^{b,e*}$}
    \IEEEauthorblockA{$^a$ School of Computer Software, College of Intelligence and Computing, Tianjin University, Tianjin, China}
    \IEEEauthorblockA{$^b$ School of Software, Northwestern Polytechnical University, Shaanxi, China}
    \IEEEauthorblockA{$^c$ Department of Computer Science and Information Technology, La Trobe University, Melbourne, Australia}
    \IEEEauthorblockA{$^d$ Department of Computer Science, School of Engineering, Shantou University, Guangdong, China}
    \IEEEauthorblockA{$^e$ Yangtze River Delta Research Institute of Northwestern Polytechnical University, Taicang, China}
    \IEEEauthorblockA{qgking@nwpu.edu.cn}
    \thanks{This work was supported by the National Natural Science Foundation of China [Grant No. 62222311, No. 62201460, and No. 32300527], the Fundamental Research Funds for the Central Universities, and the Basic Research Programs of Taicang [Grant No. TC2023JC22]. (Corresponding author: Qiangguo Jin)}
}

\maketitle

\begin{abstract}
Molecular subtyping of cancer is recognized as a critical and challenging upstream task for personalized therapy. Existing deep learning methods have achieved significant performance in this domain when abundant data samples are available. However, the acquisition of densely labeled samples for cancer molecular subtypes remains a significant challenge for conventional data-intensive deep learning approaches. In this work, we focus on the few-shot molecular subtype prediction problem in heterogeneous and small cancer datasets, aiming to enhance precise diagnosis and personalized treatment. We first construct a new few-shot dataset for cancer molecular subtype classification and auxiliary cancer classification, named TCGA Few-Shot, from existing publicly available datasets. To effectively leverage the relevant knowledge from both tasks, we introduce a task-specific embedding-based meta-learning framework (TSEML). TSEML leverages the synergistic strengths of a model-agnostic meta-learning (MAML) approach and a prototypical network (ProtoNet) to capture diverse and fine-grained features. Comparative experiments conducted on the TCGA Few-Shot dataset demonstrate that our TSEML framework achieves superior performance in addressing the problem of few-shot molecular subtype classification.
\end{abstract}


\section{Introduction}
Cancer is one of the deadliest diseases worldwide~\cite{ushijima2021mapping}. The classification of cancer molecular subtypes aids in a deeper understanding of the pathogenesis of cancer, which is beneficial for precise diagnosis and personalized treatment~\cite{liang2014integrative}. However, multiple pathogenetic mechanisms and heterogeneity in cancer pose significant obstacles to precise molecular subtyping~\cite{stingl2007molecular, dai2015breast}. Automatically defining and separating cancer subtypes using computer-aided techniques is highly demanded to facilitate personalized therapy and improve prognosis~\cite{chen2022cancer,liang2014integrative}.

Various algorithms have been developed to effectively identify cancer subtypes~\cite{Sarkar2023Firefly-SVM,Sun2023Automated,liu2004application,zhong2021a}. Despite these advancements, cancer molecular subtyping remains challenging, primarily due to the scarcity of samples for certain subtypes~\cite{yang2020select}. A prevalent strategy to address this small data (few-shot~\cite{yu2021few, chen2023few}) problem is integrative analysis, which aims to enlarge the dataset by amalgamating data from different experiments and platforms~\cite{li2014meta, hughey2015robust,gao2023biostd}. However, the complex nature of gene expression data in computational biology means that data aggregation from different studies can be negatively affected by batch effects and heterogeneity introduced by integrative approaches~\cite{lazar2012survey}. Thus, there is a critical need for datasets specifically designed for few-shot training, as well as for the development of novel methods that can accurately classify molecular cancer subtypes.

Recently, meta-learning~\cite{a2024advances}, also known as learning to learn, has emerged as a common few-shot learning framework. Two classical approaches in meta-learning are model-agnostic meta-learning (MAML)~\cite{finn2017model} and prototypical network (ProtoNet)~\cite{snell2017prototypical}, which demonstrate powerful ability in solving small data problems. However, these two classical methods may have limitations. Firstly, in the traditional MAML framework, meta-initialization contains useful features that can mostly be reused for new tasks, resulting in minimal task-specific adaptation. Secondly, ProtoNet has been observed to be prone to overfitting issues in gene expression data~\cite{schmiedel2019empirical}. Based on these observations, we hypothesize that (1) introducing an auxiliary task could prevent the model from overfitting on the cancer subtyping task; (2) designing a task-specific learning paradigm that explores the correlation between the auxiliary task and the primary cancer subtyping task could facilitate knowledge transfer between tasks, potentially enhancing the overall performance and robustness of the model.

To this end, we first construct an $N$-way $K$-shot multi-task dataset for cancer molecular subtype classification and auxiliary cancer classification. Second, we propose a task-specific embedding-based meta-learning (TSEML) framework designed to leverage interrelated clinical tasks, facilitating the extraction of common knowledge and benefiting each individual task. The main contributions of this paper can be summarized as follows:

\begin{itemize}

\item To address the scarcity of few-shot datasets in bioinformatics, we construct a dataset named TCGA Few-Shot, using data from The Cancer Genome Atlas (TCGA). TCGA Few-Shot serves as a benchmark for cancer molecular subtype classification and cancer classification tasks within a few-shot learning framework.

\item We introduce a novel meta-learning framework, TSEML, which integrates MAML and ProtoNet to complement each other. The TSEML leverages ProtoNet, enhancing the model's capability to test with varying numbers of classes. Additionally, TSEML capitalizes on MAML's capabilities while enabling ProtoNet to conduct task-specific embeddings, thereby augmenting the method's learning capacity.

\item Comprehensive experiments demonstrate that TSEML outperforms other few-shot methods in both cancer molecular subtype classification and cancer classification on the TCGA Few-Shot dataset.

\end{itemize}
\section{Materials and methods}
\subsection{Datasets and task generation}
\subsubsection{Datasets and preprocessing}
Due to the scarcity of few-shot learning datasets for molecular subtyping, we create a new TCGA Few-Shot dataset, derived from the TCGA~\cite{tomczak2015cancer}  datasets. This dataset can be used to benchmark various few-shot learning methods, supporting cancer subtype classification and cancer classification tasks.

We utilize data from UCSC Xena~\cite{goldman2019ucsc}, which performs a log2 transformation on normalized counts downloaded from the TCGA Data Coordinating Center~\cite{tomczak2015cancer} and apply mean normalization for each gene across all TCGA cohorts. 
Then, we download 10,459 IlluminaHiSeq pan-cancer based normalized gene expression profiles, encompassing 33 cancer types and 20,530 genes, from UCSC Xena, with normal samples removed. These 33 cancers are used for cancer classification.
For subtype classification, we select 14 cancer molecular subtypes from the 33 cancers. The remaining cancer molecular subtypes are excluded due to insufficient sample sizes. Detailed information on the selected types of cancers can be found in our GitHub repository~\footnote{https://github.com/BioMedIA-repo/TSEML.git}.
\begin{figure}
\centering
\includegraphics[scale=0.28]{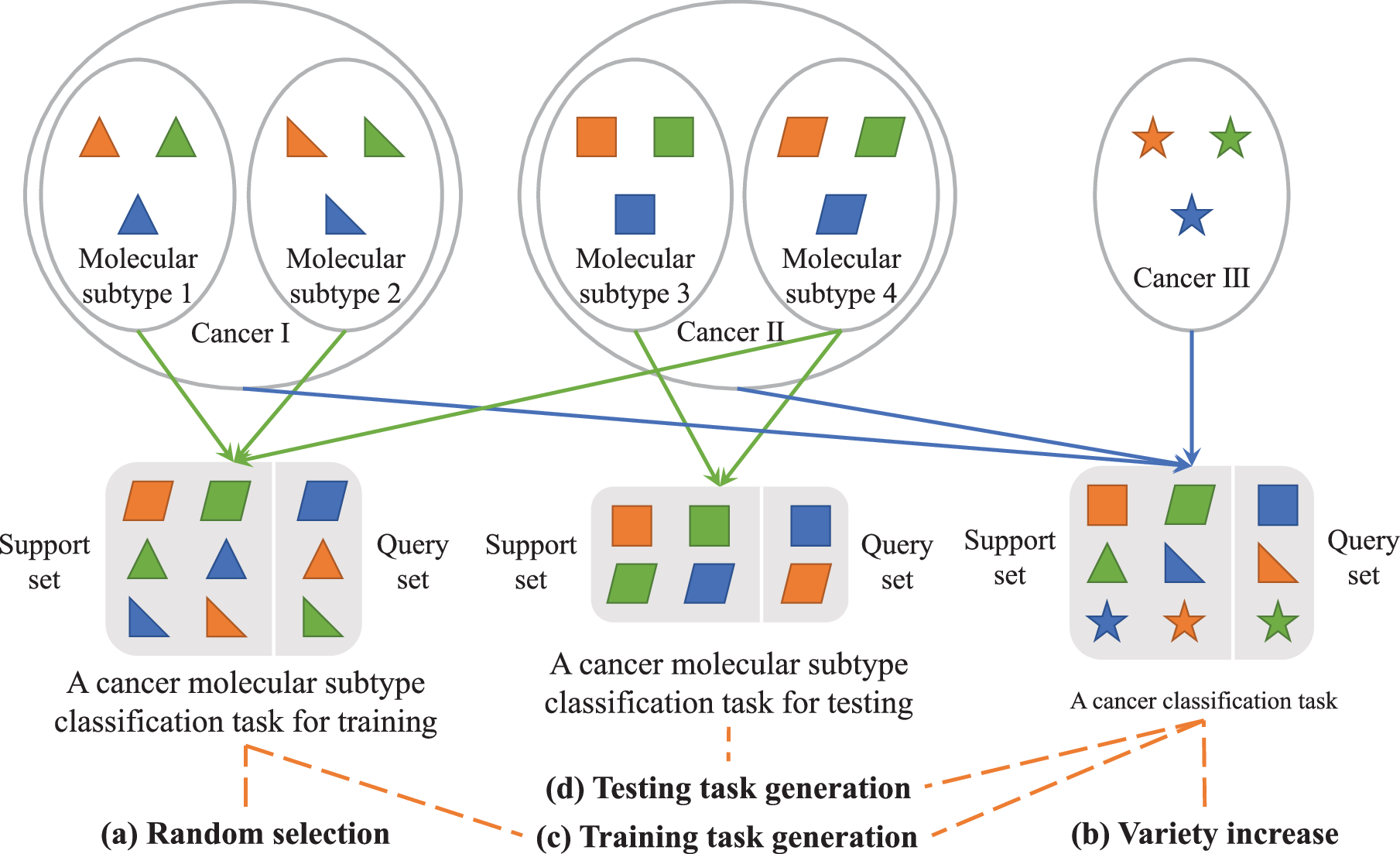}
\caption{Overview of (a) Random selection, (b) Variety increase, (c) Training task generation, and (d) Testing task generation.}
\label{fig1}
\end{figure}
\subsubsection{Few-shot learning setting}
Given a dataset $D=\left \{\left (\mathbf {x}_{k}, y_{k}\right )\right \} _{k=1}^{|D|}$, and its corresponding class set $Y=\bigcup _{k=1}^{|D|}\left \{y_{k}\right \}$. We can define an $N$-way $K$-shot task $\mathcal {T}_{i}=\left (\mathcal {S}_{i}, \mathcal {Q}_{i}\right )$ randomly sampled from $D$. The support set $\mathcal {S}_{i}$ contains $N$ classes with $K$ samples per class, while the query set $\mathcal {Q}_{i}$ includes the same $N$ classes. The number of class types $Y_{i}\subseteq Y$ in task $\mathcal {T}_{i}$ is set to be $N=\left |Y_{i}\right |$. For each $y_{i, j}\in Y_{i}(j=1,2, \ldots , N)$, we have $\mathcal {S}_{i, j}=\left \{ (\mathbf {x}, y)\mid y=y_{i, j}\right \} \subset D$, with $\left |\mathcal {S}_{i, j}\right |=K$ and $\mathcal {S}_{i}=\bigcup _{j=1}^{N}\mathcal {S}_{i, j}$. Regarding the $Q$-query setting, for every $y_{i, j}\in Y_{i}(j=1,2, \ldots , N)$, we have $\mathcal {Q}_{i, j}=\left \{ (\mathbf {x}, y)\mid y=y_{i, j}\right \} \subset \left (D-\mathcal {S}_{i, j}\right )$, $\left |\mathcal {Q}_{i, j}\right |=Q$ and $\mathcal {Q}_{i}=\bigcup _{j=1}^{N}\mathcal {Q}_{i, j}$.

\begin{figure*}
    \centering
    \includegraphics[scale=0.47]{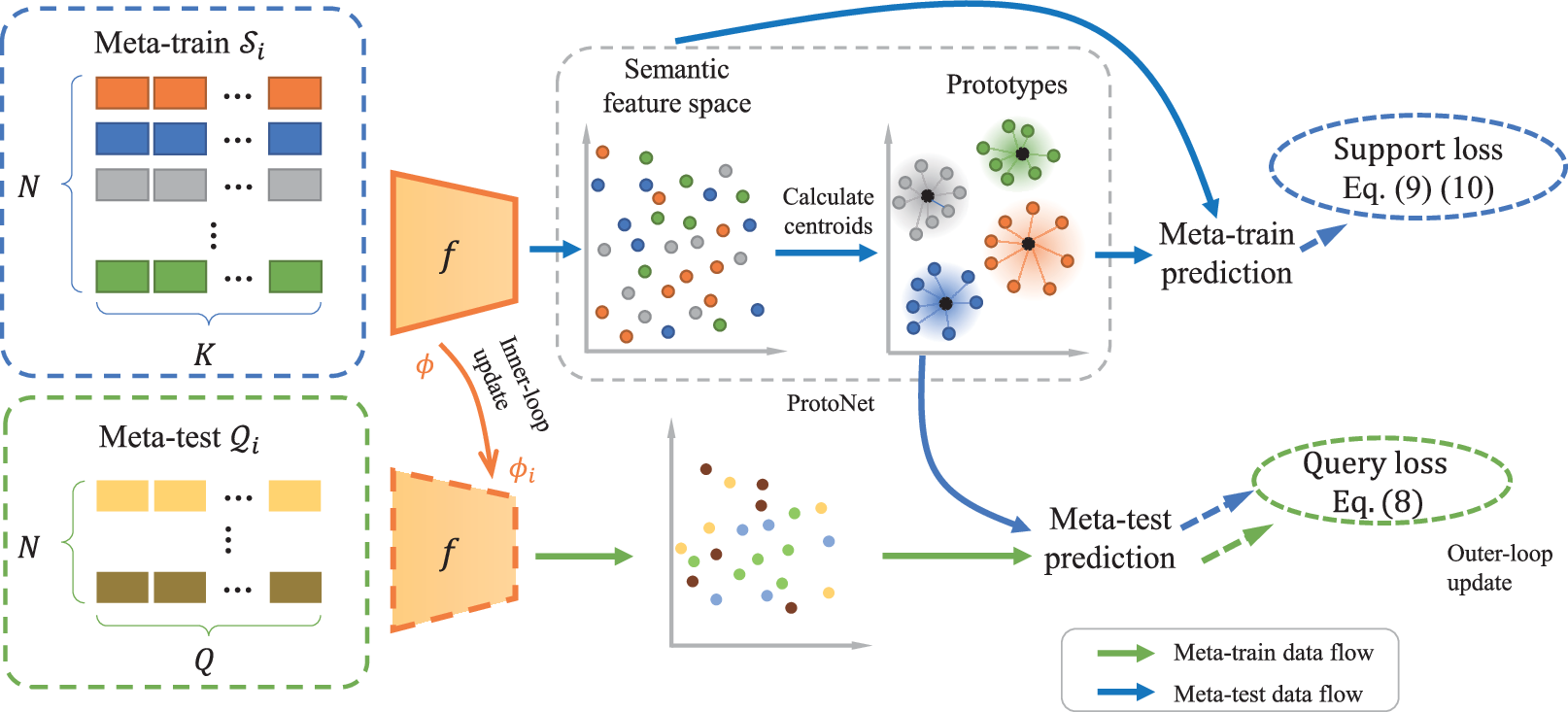}
    \caption{The training paradigm of the proposed TSEML.
    TSEML leverages the nearest feature centroid approach of ProtoNet and the gradient update inner loop of MAML to enable the model to perform task-specific embeddings for samples.}
    \label{fig2}
\end{figure*}

\subsubsection{Task generation}
To address the issue of insufficient molecular subtype samples and leverage additional knowledge from auxiliary tasks, we generate basic few-shot tasks and propose the TCGA Few-Shot dataset using the following steps (as depicted in Fig.~\ref{fig1}):

\begin{itemize}
    \item \textbf{Random selection.} Initially, we increase the diversity of training tasks by randomly sampling different cancer molecular subtypes to compose the training dataset.
    \item \textbf{Variety increase.} To further increase the variety of tasks and enhance the learning of potential knowledge, we select cancer samples, excluding those with molecular subtypes in IlluminaHiSeq, to form new tasks. This strategy facilitates the inclusion of multiple cancer molecular subtypes from a single cancer type, thereby enriching the diversity of knowledge available for more effective subtyping.
    \item \textbf{Training task generation.} We set the probability of sampling cancers for classification and molecular subtypes of a single cancer for subtyping at an equal rate. If the number of molecular subtypes is insufficient to meet expectations, additional molecular subtypes from other cancers will be sampled until the $N$-way setting requirement is satisfied, thereby forming a comprehensive cancer molecular subtype classification task.
    \item \textbf{Testing task generation.} We generate tasks that encompass all classes involved in both cancer molecular subtype classification and cancer classification
\end{itemize}

\subsection{Proposed TSEML for few-shot classification}
The TSEML is designed towards a two-fold goal: to enhance the flexibility of MAML, while also exploring various tasks with interrelated information in ProtoNet. The architecture of TSEML is depicted in Fig.~\ref{fig2}.

\subsubsection{MAML framework}
MAML~\cite{finn2017model} is a representative approach of optimization-based meta-learning, which consists of outer and inner gradient update loops. The outer loop handles the gradient update to find initial model parameters suitable for all tasks, while the inner loop optimizes the model for specific tasks.

The classification model in MAML is denoted as $f_{\theta}$, where $\theta$ represents the model parameters. Given the task distribution $p(\mathcal T)$ and the task-level learning rate $\alpha$, the optimization objective of MAML can be formulated as follows:
\begin{eqnarray}
\label{eq2}
    \theta ^{*}=\arg \min _{\theta }\mathbb {E}_{\mathcal {T}_{i}\sim p(\mathcal {T})}\frac {1}{\left |\mathcal {Q}_{i}\right |}\sum _{\left (\mathbf {x}^{q}, y^{q}\right )\in \mathcal {Q}_{i}}\mathcal {L}\left (y^{q}, f_{\theta _{i}}\left (\mathbf {x}^{q}\right )\right ),
\end{eqnarray}
where
\begin{eqnarray}
\label{eq3}
    \theta _{i}=\theta -\alpha \nabla _{\theta } \frac{1}{\left |\mathcal {S}_{i}\right |}\sum _{\left (\mathbf {x}^{s}, y^{s}\right )\in \mathcal {S}_{i}}\mathcal {L}\left (y^{s}, f_{\theta }\left (\mathbf {x}^{s}\right )\right ).
\end{eqnarray}
Alternatively, the parameters of the model can also be updated using the support set:
\begin{eqnarray}
\label{eq4}
    \theta _{i}\leftarrow \theta _{i} -\alpha \nabla _{\theta _{i} } \frac{1}{\left |\mathcal {S}_{i}\right |}\sum _{\left (\mathbf {x}^{s}, y^{s}\right )\in \mathcal {S}_{i}}\mathcal {L}\left (y^{s}, f_{\theta _{i}}\left (\mathbf {x}^{s}\right )\right ).
\end{eqnarray}

\subsubsection{ProtoNet}
ProtoNet~\cite{snell2017prototypical} is a classical metric-based meta-learning approach. In ProtoNet, samples across various tasks are embedded into a high-level feature space. Within this space, features corresponding to the same class are close to each other, while features from different classes are distinctly separated. Classification is then performed by determining the proximity of sample features to the nearest class centroids, effectively leveraging the spatial relationships within the feature space for decision-making

Let $g_{\phi}$ denote the network body with model parameters $\phi$, which can be considered as the embedding model for feature extraction. The centroid of sample features belonging to the $j$-th class of the support set $\mathcal {S}_{i}$ is given by:
\begin{eqnarray}
\label{eq5}
    \mathbf {c}_{i, j}=\frac {1}{\left |\mathcal {S}_{i, j}\right |} \sum _{(\mathbf {x}, y)\in \mathcal {S}_{i, j}} g_{\phi }(\mathbf {x}).
\end{eqnarray}
Let $d$ denote the distance function, for every $\left(\mathbf {x}^{q}, y^{q}\right ) \in \mathcal {Q}_{i}$, we use $-d\left (g_{\phi }\left (\mathbf {x}^{q}\right ), \mathbf {c}_{i, j}\right )$ to estimate the distance between the prediction of $\mathbf {x}^{q}$ and the class $y_{i, j}$. The prediction vector $\mathbf {z}^{q}$ of $\mathbf {x}^{q}$ is calculated as follows:
\begin{eqnarray}
\label{eq6}
    \mathbf {z}^{q}=\left [-d\left (g_{\phi }\left (\mathbf {x}^{q}\right ), \mathbf {c}_{i, 1}\right ), -d\left (g_{\phi }\left (\mathbf {x}^{q}\right ), \mathbf {c}_{i, 2}\right ), \ldots,\notag\right. \\ \left. -d\left (g_{\phi }\left (\mathbf {x}^{q}\right ), \mathbf {c}_{i, N}\right )\right ].
\end{eqnarray}
We denote the entire process of computing $\mathbf {z}^{q}$ by Eq~(\ref{eq5}) and~(\ref{eq6}) as the function $h_{\phi}$, i.e.,
\begin{eqnarray}
\label{eq7}
    h_{\phi }\left (\mathbf {x}^{q}, \mathcal {S}_{i}\right)= \mathbf {z}^{q}.
\end{eqnarray}
To this end, the optimization objective of ProtoNet is defined as follows:
\begin{eqnarray}
\label{eq8}
    \phi ^{*}=\arg \min _{\phi} \mathbb {E}_{\mathcal {T}_{i}\sim p(\mathcal {T})}\frac {1}{\left |\mathcal {Q}_{i}\right |}\sum _{\left (\mathbf {x}^{q}, y^{q}\right )\in \mathcal {Q}_{i}} \mathcal {L}\left (y^{q}, h_{\phi }\left (\mathbf {x}^{q}, \mathcal {S}_{i}\right )\right ).
\end{eqnarray}

\subsubsection{TSEML}
The traditional MAML and ProtoNet suffer from several limitations: (1) The network head of MAML limits the number of classes for classification, and exacerbates the negative impact of low representative features with introduced uncertainties. (2) ProtoNet fails in modeling relevant knowledge, and changes across different classification tasks.


To alleviate these limitations, we propose the TSEML framework, as shown in Fig.~\ref{fig2}. TSEML employs the classification method based on the nearest feature centroid of ProtoNet, which not only reduces the uncertainty of the method but also allows the model to handle classification tasks with an arbitrary number of classes. Additionally, TSEML utilizes the gradient update inner loop of MAML to perform task-specific embeddings for samples. By such, the learning ability of TSEML is enhanced significantly. The optimization objective of TSEML is calculated as follows:
\begin{eqnarray}
\label{eq9}
    \phi ^{*}=\arg \min _{\phi } \mathbb {E}_{\mathcal {T}_{i}\sim p(\mathcal {T})}\frac {1}{\left |\mathcal {Q}_{i}\right |}\sum _{\left (\mathbf {x}^{q}, y^{q}\right )\in \mathcal {Q}_{i}}\mathcal {L}\left (y^{q}, h_{\phi _{i}}\left (\mathbf {x}^{q}, \mathcal {S}_{i}\right )\right ),
\end{eqnarray}
where
\begin{eqnarray}
\label{eq10}
    \phi _{i}=\phi -\alpha \nabla _{\phi }\frac {1}{\left |\mathcal {S}_{i}\right |}\sum _{\left (\mathbf {x}^{s}, y^{s}\right )\in \mathcal {S}_{i}}\mathcal {L}\left (y^{s}, h_{\phi }\left (\mathbf {x}^{s}, \mathcal {S}_{i}\right )\right ).
\end{eqnarray}
As with MAML, we can also perform multiple updates using the support set:
\begin{eqnarray}
\label{eq11}
    \phi _{i}\leftarrow \phi _{i}-\alpha \nabla _{\phi _{i}}\frac {1}{\left |\mathcal {S}_{i}\right |}\sum _{\left(\mathbf {x}^{s}, y^{s}\right)\in \mathcal {S}_{i}}\mathcal {L}\left (y^{s}, h_{\phi _{i}}\left (\mathbf {x}^{s}, \mathcal {S}_{i}\right )\right ).
\end{eqnarray}

\subsubsection{Loss function}
We use the cross-entropy as the $N$-way setting loss function:
\begin{eqnarray}
\label{eq1}
    \mathcal {L}(y, \mathbf {z})=\ln {\sum _{j=1}^{N}\exp {(\mathbf {z}[j])}}-\mathbf {z}[l(y)],
\end{eqnarray}
where $l(y)$ is the label corresponding to $y$, $\mathbf {z}$ is the output vector of the classifier, and $\mathbf {z}[j]$ represents the $j$-th element in $\mathbf {z}$.

\section{Results}
\subsection{Experimental settings}

\subsubsection{Network architecture and hyper-parameters}
Due to data scarcity and overfitting issues, we adopt a simple 1D-CNN architecture as our classification model, following the approach of Mostavi et~al.~\cite{mostavi2020convolutional}. Our method is implemented in PyTorch using an NVIDIA RTX 3090 graphic card. During the training phase of TSEML, we set the batch size of tasks to 10 and the meta-level learning rate to $1\times 10^{-4}$. TSEML is updated on the support set 5 times for training and 10 times for testing. We conduct 1,000 training iterations, testing the model after every 10 training iterations. The task-level learning rate is set to 0.01. Detailed model architecture and hyperparameter settings are available in our GitHub repository.

\subsubsection{Task settings}
All methods, except those trained without tasks, are conducted under the same meta-learning training settings for fair comparisons. We utilize two meta-learning training configurations: 5-way 1-shot 15-query and 5-way 5-shot 15-query. For testing, the tasks are set to $P$-way 1-shot 1-query and $Q$-way 5-shot 5-query, where $P$ and $Q$ are equal to the number of classes in the specific task. For each classification challenge, we randomly generate 500 tasks and calculate the mean and standard deviation of the evaluation metrics.

\subsubsection{Evaluation metrics}
We conduct ten-fold cross-validation experiments on the TCGA Few-Shot dataset. The performance of the multi-class classification is assessed using a comprehensive set of evaluation metrics, including accuracy ($\mathrm{AC}$), macro-average precision ($\mathrm{PRE}_{\mathrm{m}}$), macro-average recall ($\mathrm{REC}_{\mathrm{m}}$), macro-average F1-score ($\mathrm{F1}_{\mathrm{m}}$), and macro-average area under the receiver operating characteristic curve ($\mathrm{AUC}_{\mathrm{m}}$).

\subsection{Methods for comparison}
To demonstrate the effectiveness of our TSEML, we compare it with typical models, which can be categorized into traditional deep learning based methods (VGG~\cite{ye2021vgg}, deep neural network (DNN)~\cite{Hajieskandar2023dnn}, recurrent neural network (RNN)~\cite{Thakur2023rnn} and classifier-baseline (CB)~\cite{chen2021meta}), and meta-learning based methods (meta-long short term memory (ML)~\cite{zhang2021ml}, meta-stochastic gradient descent (MS)~\cite{yan2021ms}, contextual augmented meta-learning (CAML)~\cite{israr2023caml}, meta-baseline (MB)~\cite{chen2021meta}, MAML~\cite{raghu2019rapid}, and ProtoNet~\cite{snell2017prototypical}).
    
\begin{table}
    \centering
    \caption{The average experimental results (mean $\pm$ std) of different methods on cancer molecular subtype classification tasks. The best results are highlighted in bold.}
    \renewcommand\tabcolsep{1.1pt}
    \renewcommand\arraystretch{1.1}
    \begin{tabular}{c|c|c|c|c|c}
    \hline
    \tabincell{c}{Test\\setting} & Method & $\mathrm{AC}$ (\%) & $\mathrm{PRE}_{\mathrm{m}}$ (\%) & $\mathrm{F1}_{\mathrm{m}}$ (\%) & $\mathrm{AUC}_{\mathrm{m}}$ (\%)\\ \hline
    \multirow{10}{*}{1-shot}& VGG & 27.46 ± 4.86 & 52.94 ± 3.40 & 21.36 ± 4.30 & 51.37 ± 2.88\\
    & DNN & 28.10 ± 3.71 & 70.38 ± 7.87 & 15.94 ± 2.92 & 51.84 ± 2.11\\
    & RNN & 43.06 ± 17.76 & \textbf{73.85 ± 8.31} & 32.96 ± 19.39 & 51.02 ± 2.06\\
    & CB & 48.65 ± 8.01 & 70.32 ± 5.43 & 39.70 ± 7.83 & 73.86 ± 8.44\\
    & ML & 37.67 ± 19.76 & 71.68 ± 11.73 & 27.26 ± 22.03 & 50.09 ± 2.87\\
    & MS & 28.08 ± 4.14 & 55.15 ± 3.89 & 21.00 ± 3.84 & 51.86 ± 1.88\\ 
    & CAML & 26.74 ± 4.10 & 63.91 ± 8.76 & 16.79 ± 3.60 & 50.17 ± 1.21\\ 
    & MB & 48.81 ± 7.99 & 70.33 ± 5.24 & 39.98 ± 7.88 & 75.10 ± 8.85\\ 
    & MAML & 47.85 ± 7.75 & 70.25 ± 5.33 & 38.74 ± 7.50 & 74.11 ± 8.58\\ 
    & ProtoNet & 49.15 ± 8.16 & 71.40 ± 4.90 & 40.11 ± 8.07 & 75.77 ± 8.68\\ 
    & TSEML & \textbf{51.08 ± 9.17} & 72.83 ± 5.90 & \textbf{41.91 ± 9.26} & \textbf{77.82 ± 9.17}\\  \hline
    \multirow{10}{*}{5-shot} & VGG & 26.40 ± 3.89 & 34.55 ± 15.31 & 22.70 ± 4.54 & 50.31 ± 1.14\\ 
    & DNN & 28.47 ± 3.84 & 65.75 ± 6.94 & 17.77 ± 4.16 & 51.69 ± 2.15\\
    & RNN & 51.81 ± 27.56 & 72.78 ± 16.52 & 43.98 ± 32.02 & 50.11 ± 1.44\\
    & CB & 63.63 ± 12.96 & 67.25 ± 13.40 & 62.51 ± 13.25 & 83.43 ± 9.81\\ 
    & ML & 31.76 ± 8.93 & 49.33 ± 18.32 & 25.60 ± 10.81 & 50.64 ± 1.07\\ 
    & MS & 26.74 ± 3.63 & 43.63 ± 14.81 & 20.95 ± 5.92 & 50.40 ± 1.37\\ 
    & CAML & 26.99 ± 4.41 & 54.88 ± 13.10 & 17.93 ± 5.58 & 51.20 ± 1.48\\ 
    & MB & 65.06 ± 12.52 & 68.73 ± 13.10 & 63.97 ± 12.83 & 85.89 ± 10.18\\
    & MAML & 60.94 ± 11.78 & 64.83 ± 12.07 & 59.36 ± 11.99 & 83.15 ± 9.96\\ 
    & ProtoNet & 64.51 ± 13.08 & 68.11 ± 13.69 & 63.30 ± 13.44 & 85.44 ± 10.66\\ 
    & TSEML & \textbf{70.84 ± 13.32} & \textbf{74.26 ± 13.43} & \textbf{69.86 ± 13.73} & \textbf{89.34 ± 9.80}\\ \hline
    \end{tabular}
    \label{table1}
    \end{table}

\begin{table}
    \centering
    \caption{The average experimental results (mean $\pm$ std) of different methods on cancer classification tasks. The best results are highlighted in bold.}
    \renewcommand\tabcolsep{1.1pt}
    \renewcommand\arraystretch{1.1}
    \begin{tabular}{c|c|c|c|c|c}
    \hline
    \tabincell{c}{Test\\setting} & Method & $\mathrm{AC}$ (\%) & $\mathrm{PRE}_{\mathrm{m}}$ (\%) & $\mathrm{F1}_{\mathrm{m}}$ (\%) & $\mathrm{AUC}_{\mathrm{m}}$ (\%)\\ \hline
    \multirow{10}{*}{1-shot} & VGG & 32.78 ± 6.09 & 53.25 ± 5.21 & 25.93 ± 5.14 & 49.67 ± 2.69\\ 
    & DNN & 33.58 ± 3.38 & 61.41 ± 8.81 & 23.88 ± 4.06 & 50.67 ± 3.67\\ 
    & RNN & 54.74 ± 22.19 & 71.85 ± 14.16 & 47.82 ± 25.11 & 51.83 ± 3.30\\ 
    & CB & 93.02 ± 5.89 & 96.17 ± 3.46 & 90.97 ± 7.45 & 97.94 ± 2.44\\ 
    & ML & 45.18 ± 13.76 & 64.22 ± 10.82 & 37.45 ± 13.91 & 52.63 ± 2.39\\ 
    & MS & 33.34 ± 3.85 & 56.48 ± 2.89 & 25.83 ± 4.38 & 52.49 ± 2.41\\
    & CAML & 31.67 ± 4.09 & 54.45 ± 6.25 & 24.50 ± 4.86 & 50.51 ± 1.18\\ 
    & MB & \textbf{95.61 ± 4.99} & \textbf{97.59 ± 2.92} & \textbf{94.31 ± 6.33} & \textbf{99.25 ± 1.24}\\ 
    & MAML & 92.93 ± 6.27 & 96.17 ± 3.63 & 90.83 ± 7.96 & 98.57 ± 1.88\\
    & ProtoNet & 94.34 ± 6.10 & 96.88 ± 3.62 & 92.69 ± 7.70 & 99.00 ± 1.75\\ 
    & TSEML & 95.07 ± 5.38 & 97.34 ± 3.11 & 93.59 ± 6.86 & \textbf{99.25 ± 1.45}\\  \hline
    \multirow{10}{*}{5-shot}& VGG & 31.39 ± 3.06 & 36.02 ± 6.98 & 28.01 ± 2.89 & 50.47 ± 1.10\\
    & DNN & 34.78 ± 4.96 & 55.77 ± 6.87 & 27.54 ± 4.19 & 53.03 ± 2.86\\ 
    & RNN & 57.73 ± 24.80 & 65.76 ± 20.63 & 53.71 ± 27.08 & 50.83 ± 2.06\\ 
    & CB & 97.32 ± 2.83 & 97.76 ± 2.40 & 97.22 ± 3.01 & 99.22 ± 1.46\\
    & ML & 42.56 ± 8.15 & 46.89 ± 6.33 & 38.82 ± 8.82 & 52.90 ± 1.72\\
    & MS & 32.54 ± 5.40 & 46.17 ± 13.84 & 27.39 ± 7.58 & 51.39 ± 3.43\\
    & CAML & 34.33 ± 4.91 & 42.39 ± 9.36 & 29.33 ± 5.15 & 52.57 ± 2.09\\ 
    & MB & 98.14 ± 2.42 & 98.45 ± 2.01 & 98.08 ± 2.56 & 99.83 ± 0.41\\
    & MAML & 96.99 ± 3.52 & 97.50 ± 2.96 & 96.89 ± 3.70 & 99.55 ± 0.76\\ 
    & ProtoNet & 97.48 ± 3.63 & 97.91 ± 3.06 & 97.36 ± 3.92 & 99.65 ± 0.81\\ 
    & TSEML & \textbf{98.41 ± 2.30} & \textbf{98.67 ± 1.94} & \textbf{98.37 ± 2.39} & \textbf{99.86 ± 0.36}\\  \hline
    \end{tabular}
    \label{table2}
    \end{table}

\begin{figure*}
    \includegraphics[scale=0.68]{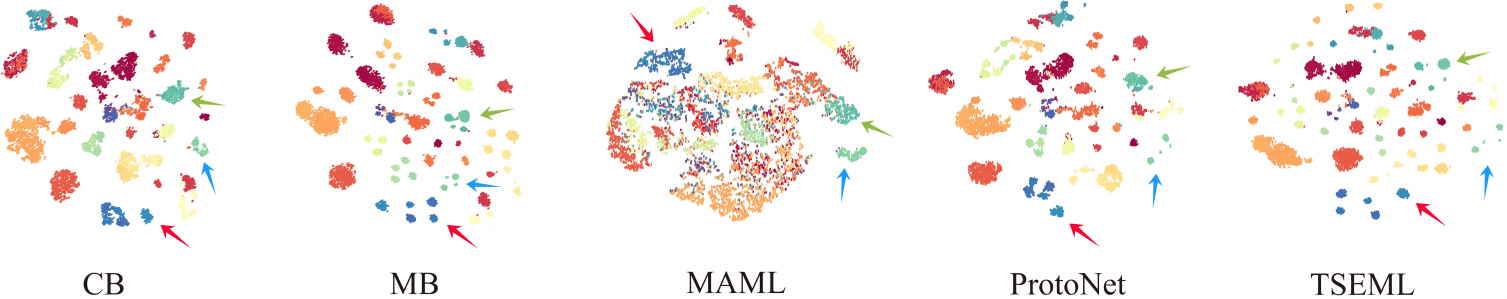}
    \caption{ Visual interpretation of high-level embeddings learned by different methods on TCGA Few-Shot. The red, blue, and green arrows highlight the ability of our TSEML to discriminate specific subtypes.}
    \label{figS2}
\end{figure*}

\subsection{Comparison results}
\subsubsection{Molecular subtype classification}
As shown in Table~\ref{table1}, our method, TSEML, consistently outperform other approaches in both 1-shot and 5-shot settings. 

In the 1-shot setting, TSEML achieves the highest $\mathrm{AUC}_{\mathrm{m}}$ of 77.82\%, showing a notable improvement over the closest competitor, ProtoNet, which records an $\mathrm{AUC}_{\mathrm{m}}$ of 75.67\%. Furthermore, TSEML attains a $\mathrm{PRE}_{\mathrm{m}}$ of 72.83\% and an $\mathrm{F1}_{\mathrm{m}}$ of 41.91\%, indicating the robustness of our methods in handling minimal and unbalanced data scenarios.
In the 5-shot setting, the performance improvement of our methods is even more significant.  TSEML achieves the best results across all metrics

These results underscore the effectiveness of our methods in enhancing the precision and reliability of cancer molecular subtype classification in few-shot learning scenarios.

\subsubsection{Cancer classification}
Our experimental analysis on the auxiliary cancer classification tasks, as shown in Table~\ref{table2}, presents a comparative study of various methods.

In the 1-shot setting, our TSEML achieves the best $\mathrm{AUC}_{\mathrm{m}}$ of 99.25\%. 
In the 5-shot setting, TSEML further consolidates its lead by posting the highest scores in all metrics.

In summary, our first finding is that the TSEML framework outperforms other methods by a large margin on all tasks except cancer classification. We explain this result as the reason that the diversity of this dataset for cancer classification is limited, and the classical methods have reached the upper-bound performance on the dataset. However, TSEML demonstrates its capability to achieve precise predictions, highlighting its robust learning ability. As for cancer molecular subtype classification, our TSEML shows its powerful ability in learning highly representative embedding by a larger support set for gradient updates.

\subsection{Ablation study for distance metric functions}
To evaluate the impact of different distance metric functions within TSEML, we conduct ablation experiments using Euclidean distance (TSEML) and cosine distance (TSEML-C).

\textbf{Molecular subtype classification.} 
In the 1-shot setting, TSEML achieves slightly higher metrics overall compared to TSEML-C, as shown in Table~\ref{table1}, with an $\mathrm{AC}$ of 51.08\% versus 51.72\%, and an $\mathrm{AUC}_{\mathrm{m}}$ of 77.82\% compared to 77.17\%. TSEML also records a higher $\mathrm{PRE}_{\mathrm{m}}$ at 72.83\% compared to 72.26\%, though TSEML-C achieves a higher $\mathrm{F1}_{\mathrm{m}}$ at 48.2\% compared to 41.91\%.

\begin{table}[]
    \centering
    \caption{The average experimental results (mean $\pm$ std) of TSEML framework with different distance metric functions. }
    \renewcommand\tabcolsep{1.1pt}
    \renewcommand\arraystretch{1.1}
    \begin{tabular}{c|c|c|c|c|c|c}
    \hline
    Task &  \tabincell{c}{Test\\setting} & Method &  $\mathrm{AC}$ (\%) & $\mathrm{PRE}_{\mathrm{m}}$ (\%) & $\mathrm{F1}_{\mathrm{m}}$ (\%) &  $\mathrm{AUC}_{\mathrm{m}}$ (\%) \\  \hline
    \multirow{8}{*}{\tabincell{c}{CMSC}} & \multirow{4}{*}{1-shot} & TSEML-X & & & &  \\ 
     &  & TSEML-X & & & &  \\
     & & TSEML-C & 51.72 ± 9.32 & 72.26 ± 6.00 & 42.82 ± 9.41 & 76.30 ± 9.14\\ 
     & & TSEML & 51.08 ± 9.17 & 72.83 ± 5.90 & 41.91 ± 9.26 & 77.82 ± 9.17\\ \cline{2-7}
     & \multirow{4}{*}{5-shot} & TSEML-X &  & & &  \\
     &  &  & & & & \\
     & & TSEML-C & 68.06 ± 13.29 & 71.44 ± 13.48 & 66.84 ± 13.72 & 87.10 ± 9.99 \\
     & & TSEML & 70.84 ± 13.32 & 74.26 ± 13.43 & 69.86 ± 13.73 & 89.34 ± 9.80 \\ \hline
     \multirow{8}{*}{\tabincell{c}{CC}} & \multirow{4}{*}{1-shot} & TSEML-X & & & &  \\ 
     &  &  TSEML-Y & & & &\\
     &  & TSEML-C & 95.74 ± 5.46 & 97.66 ± 3.20 & 94.47 ± 6.94 & 98.97 ± 1.88\\ 
     &  & TSEML & 95.07 ± 5.38 & 97.34 ± 3.11 & 93.59 ± 6.86 & 99.25 ± 1.45\\ \cline{2-7}
     & \multirow{4}{*}{5-shot} & TSEML-X &  & &  &  \\
     &  & TSEML-Y & & & &\\
     & & TSEML-C & 98.15 ± 3.01 & 98.48 ± 2.46 & 98.04 ± 3.30 & 99.54 ± 1.25\\ 
     & & TSEML & 98.41 ± 2.30 & 98.67 ± 1.94 & 98.37 ± 2.39 & 99.86 ± 0.36\\  \hline
    \end{tabular}
    \end{table}

In the 5-shot setting, TSEML continues to outperform TSEML-C, achieving an $\mathrm{AC}$ of 70.84\% versus 68.06\%, and an $\mathrm{AUC}_{\mathrm{m}}$ of 89.34\% compared to 87.10\%. TSEML also maintains a higher $\mathrm{PRE}_{\mathrm{m}}$ (74.26\% versus 71.44\%) and $\mathrm{F1}_{\mathrm{m}}$ (69.86\% versus 66.84\%). These results indicate that while both methods are highly effective, TSEML generally provides slightly better performance, particularly in terms of precision and overall accuracy.

\textbf{Cancer classification.} The performance variances between TSEML and TSEML-C on the cancer classification task, as shown in Table~\ref{table2}, are minimal, indicating that both methods are nearly equally effective.

Our finding is that different distance metric functions have a significant impact on various tasks. Specifically, molecular subtype classification is notably sensitive to the choice of metric functions. This sensitivity can be attributed to the complexity of the molecular subtype classification task, where the Euclidean distance tends to be more effective in capturing the inherent relationships between tasks.

\subsection{t-SNE visualization results}
We employ t-SNE~\cite{van2008visualizing} to visualize the distributions of samples within the high-level semantic latent feature embeddings constructed by the TSEML models under the 1-shot setting. The resulting visualization, as illustrated in Fig.~\ref{figS2}, reveals that TSEML successfully clusters samples in the high-level semantic space.

\section{Conclusion}
In this study, we introduce a task-specific embedding-based meta-learning framework (TSEML), specifically designed for the few-shot classification of cancer molecular subtypes. The proposed approach synergistically combines the strengths of the model-agnostic meta-learning framework and the prototypical network. TSEML is capable of extracting shareable knowledge from interconnected cancer molecular subtype classification tasks, effectively learning from a limited number of samples to enhance the network's task-specific learning capabilities. Additionally, we have developed a few-shot benchmark dataset, TCGA Few-Shot, specifically for cancer molecular subtype classification. Comparative experiments on the TCGA Few-Shot dataset demonstrate that TSEML gains superiority in addressing the challenges of few-shot classification of cancer molecular subtypes.

\bibliographystyle{elsarticle-num.bst}
\bibliography{references}

\end{document}